# Modeling Opinion Dynamics: Ranking Algorithms on Heterogeneous Populations

Ivan V. Kozitsin
*Laboratory of Active Systems*
*ICS RAS*
Moscow, Russia
*Department of Higher Mathematics*
*Moscow Institute of Physics and Technology*
Dolgoprudny, Russia
kozitsin.ivan@mail.ru

*Abstract*—The heterogeneity of the influence processes is an important feature of social systems: how we perceive social influence and how we influence other individuals is heavily influenced by our opinion and non-opinion attributes. The latter include demographic, cultural, and structural (how we are embedded in social networks) characteristics. Furthermore, the results of the influence processes may also depend on how similar the interacting individuals are in terms of their features. This paper addresses this issue and elaborates on an agent-based model that is sensitive to the individual characteristics, both opinion and non-opinion ones. The model is fortified with a ranking algorithm that mimics the ranking algorithms widely adopted in real-world online social networks. For the resulting model, I elaborate a mean-field approximation that describes the behavior of the model at the macroscopic level via an autonomous system of ordinary differential equations. The properties of this system are thoroughly studied.

## I. INTRODUCTION

During recent decades, the issue of studying opinion formation models (aka social influence models) has become an independent interdisciplinary research direction that attracts attention of scholars from mathematics, physics, control theory, sociology, and social psychology. Opinion formation models explain how our views evolve in time and how influence that comes from our peers, mass media, and the Internet contributes to these processes. Understanding the mechanisms behind opinion formation is extremely important because it helps contain the spread of fake news [1] and convince citizens to follow government recommendations during the COVID-19 pandemic [2].

The classical opinion formation models that appeared in the twentieth century were usually built upon one specific local update rule [3], [4]. To date, a large number of update rules have been proposed that generate different opinion dynamics on both microscopic and macroscopic levels [5]. The huge scope of opinion dynamics models is continuously replenished. Many of these additions are of an unsystematic character [6], [7]. In this regard, the current trend in this research direction is to revise the current models to obtain the most reliable micro mechanism of opinion formation [5] and to develop generalized or unifying models (or, as they are sometimes called, frameworks) that can include or approximate several update rules at once [8], [9]. Such frameworks are of great importance because they allow comparing different opinion formation mechanisms and corresponding models built upon these mechanisms within one single mathematical frame. This issue allows scholars to bring order to the field of social influence models [10].

In the recently published paper [11], one such unifying model was proposed. In this model (hereafter – the SCARDO model, abbreviated from Stochastic Conditional ARranged Discrete Opinions), agents update their opinions stochastically, through successive pairwise interactions (if necessary, restricted by structural constraint). A special 3-D matrix stipulates the outcomes of such interactions via a probability distribution that is a function of the interacting agents' opinions. Although the model may give rise to nontrivial complex maps of opinion changes that can approximate real social dynamics quite well [11], the model assumes that all agents follow *on average* the same opinion formation mechanism. However, in real life, individual behavior may vary across demographic, cultural, and educational lines [12]. Furthermore, in online networks, special malicious accounts operate whose influence on ordinary (native) users may have a completely different effect, compared to the influence caused by the native users themselves [13]. In other words, real influence processes are heterogenous by their very nature.

In this paper, I advance the SCARDO model to account for this heterogeneity. To achieve this purpose, I assume that the outcomes of the influence events depend not only on the interacting agents' opinions but also on their non-opinion attributes. Next, I implement a ranking algorithm into the advanced SCARDO model. The algorithm broadly utilizes information on both opinion and non-opinion attributes of agents while deciding whether a given pair of agents will communicate or not. The resulting model is enriched with a mean-field approximation that describes the behavior of the model at the macroscopic level via an autonomous system of ordinary differential equations under the assumptions that the number of agents is large and the agents communicate via a complete social graph. Importantly, this approximation accounts for the effect of the ranking algorithm. The properties of the underlying system of differential equations are thoroughly studied.

The research is supported by a grant from the Russian Science Foundation (project no. 22- 71-00075).

## II. METHODS

### A. The SCARDO Model

In this subsection, I briefly introduce the SCARDO model [11]. The model investigates how $N$ agents communicate with each other via a social network $G = (V, E)$ that defines social connections between individuals, with $V$ standing for agent vertices and $E$ describing edges. Opinion dynamics are organized as successive local opinion updates. The one-to-one protocol is implemented whereby at each iteration $t$ a randomly chosen agent $i$ (influence recipient) interacts with one of their neighbors in the network – say, with an agent $j$ (influence donor), which is also selected at random (both the selection procedures are carried out according to the uniform probability distribution). As a result of the influence event, the agent $i$ may (or may not) change their opinion $o_i(t) \in X$, where the set $X$ represents a discrete opinion space:

$$X = \{x_1, \dots, x_m\}.$$

In that space, opinion values $x_1, \dots, x_m$ symbolize various positions regarding a controversial issue [14]. These positions could be arranged, if necessary:

$$x_1 \prec \cdots \prec x_m.$$

By introducing an order on $X$, we acquire an opportunity to say that this or that opinion is more left or, say, more radical, than a different one.

Let us assume that $i$'s and $j$'s opinions are $o_i(t) = x_s$ and $o_j(t) = x_l$ for some $s \in \{1, \dots, m\}$ and $l \in \{1, \dots, m\}$. The agent $i$'s opinion is changed stochastically, according to a distribution $P_{s,l} = \{p_{s,l,1}, \dots, p_{s,l,m}\}$, where:

$$p_{s,l,1} + \cdots + p_{s,l,m} = 1.$$

In the distribution $P_{s,l}$, the probability $p_{s,l,k}$ is the conditional probability of taking the opinion $x_k$ after receiving influence from the opinion $x_l$ given the current opinion is $x_s$:

$$p_{s,l,k} = \Pr\bigl(o_i(t+1) = x_k \mid o_i(t) = x_s, o_j(t) = x_l\bigr).$$

For a fixed $s$, we can build up a row-stochastic matrix that describes how the agents with an opinion $x_s$ react to peer influence:

$$P_s = \begin{bmatrix} P_{s,1} \\ \dots \\ P_{s,m} \end{bmatrix} = \begin{bmatrix} p_{s,1,1} & \dots & p_{s,1,m} \\ \dots & \dots & \dots \\ p_{s,m,1} & \dots & p_{s,m,m} \end{bmatrix} \in \mathbb{R}^{m \times m}.$$

Collected together, the matrices $P_1, \dots, P_m$ characterize exhaustively the map of social influence in the model. For convenience, this 3-D construction is denoted as the transition matrix $P$:

$$P = [P_1, \dots, P_m] \in \mathbb{R}^{m \times m \times m}.$$

The transition matrix is a key ingredient of the model. Its organization makes it possible to reproduce a broad variety of microscopic influence mechanisms (assimilative influence, bounded confidence, repulsive influence) and the opinion formation models built upon these mechanisms – see [11] for details.

### B. Accounting for heterogeneous populations

The seminal paper [11] utilizes the assumption that all agents act on average equally while being influenced by the same opinion. However, it is not a realistic representation of real-world processes. In real life, different individuals may display various levels of social power and influence susceptibility. No doubt, the way how we influence other individuals and how we are influenced by our peers depends on our characteristics. Examples: our age (younger individuals tend to be more vulnerable towards peer influence), gender (empirical studies suggest that women socialize better), education level (more educated people are more resistant to peer influence), how we perceive ourselves in society (for example, individuals having more followers may think that they are highly respected and thus their self-appraisals may feature higher values), etc. [10], [15]. Furthermore, empirical studies show that the more non-opinion features two individuals share, the more likely that they are to agree [12]. It is also worth pointing out that in the online environment, not only native users exist, but also special accounts that may pursue malicious purposes – so-called online bots. Instead of ordinary users, these accounts tend to act *strategically*. Because of this issue, their actions exert a fair effect on the social system [13]. In this regard, it is essential to distinguish between these two types of accounts while modeling online ecosystems.

To account for these observations, I suggest modifying the SCARDO model as follows. Assume that each agent is characterized by the following $L$ attributes: $X^1, \dots, X^L$. For the sake of simplicity, we will assume that opinion is the first attribute. Other $L - 1$ attributes may stand for both static and dynamic characteristics, such as age, gender, the number of friends (followees/followers), etc. Each of these attributes may take a limited number of values: $X^r = \{x_1^r, \dots, x_{m_r}^r\}$ whereby, again, $x_1^r, \dots, x_{m_r}^r$ may be arranged (for example, in the case of age) or not (if we speak, say, about gender). I assume that all characteristics are discrete. This assumption should not lead to any problems because each continuous quantity may be discretized in some way (as it usually happens in sociological surveys). Within this encoding strategy, each agent is characterized by a cortege of length $L$

$$\Bigl( \underbrace{x_{i_1}^1}_{\text{opinion}} , \underbrace{x_{i_2}^2, \dots, x_{i_L}^L}_{\text{other attributes}} \Bigr),$$

where $i_1 \in \{1, \dots, m_1\}, i_2 \in \{1, \dots, m_2\}, \dots, i_L \in \{1, \dots, m_L\}$. Overall, there are

$$M = m_1 * \prod_{j=2}^{L} m_j$$

possible cortege combinations. For convenience, I denote these combinations as $z_1, \dots, z_M$. I also believe that variables $z_1, \dots, z_{\prod_{j=2}^{L} m_j}$ signify corteges with opinion $x_1^1$, variables $z_{\prod_{j=2}^{L} m_j+1}, \dots, z_{2*\prod_{j=2}^{L} m_j}$ signify corteges with opinion $x_2^1$, and so one. Finally, $z_{(m_1-1)*\prod_{j=2}^{L} m_j}, \dots, z_{m_1*\prod_{j=2}^{L} m_j}$ denote corteges with opinion $x_{m_1}^1$. The set of all possible corteges is denoted as $Z = \{z_1, \ldots, z_M\}$.

From this moment, I assume that if two agents $i$ and $j$ communicate ($i$ is a recipient, $j$ is a donor), then $i$'s opinion is determined according to a distribution, which is a function of the recipient's and donor's corteges. In this case, to define all possible interactions (recall that there are $M^2$ possible combinations of cortege pairs) as well as all possible outcomes ($m_1$) one needs $M^2 * m_1$ conditional probabilities. However, to make analytical computations feasible, I apply a more complicated encoding strategy: I use $M^3$ variables to determine opinion dynamics. That is, when agents with corteges $z_s$ (recipient) and $z_l$ (donor) communicate, the outcome of the interaction is driven by a probability distribution $P_{s,l}^a = \{p_{s,l,1}^a, \ldots, p_{s,l,M}^a\}$, where $p_{s,l,k}^a$ stands for the conditional probability of making a cortege change $z_s \to z_k$ after being influenced by the agent with cortege $z_l$. As in the seminal model, I use the quantities $p_{s,l,k}^a$ to organize the matrix $P^a = [P_1^a, \ldots, P_m^a] \in \mathbb{R}^{M \times M \times M}$, where

$$P_s^a = \begin{bmatrix} p_{s,1,1}^a & \ldots & p_{s,1,m}^a \\ \ldots & \ldots & \ldots \\ p_{s,m,1}^a & \ldots & p_{s,m,m}^a \end{bmatrix} \in \mathbb{R}^{M \times M}.$$

The matrix $P^a$ will be referred to as the *augmented* transition matrix hereafter. It is worth pointing out that in the case where some of the individual characteristics are static, many of the elements in the augmented transition matrix will be equal to zero. Let consider the following illustrative example.

*Example 1. Let us assume that there are two agent attributes: opinion and, say, age denoted by $X^1 = \{-1, 1\}$ (a binary spin-like opinion space) and $X^2 = \{a, b\}$ (only two age groups are emphasized – $a$ and $b$). In this case, there are $M = 4$ possible corteges:*

$$z_1 = (-1, a), z_2 = (-1, b), z_3 = (1, a), z_4 = (1, b).$$

*As such, the augmented transition matrix should include $M^3 = 64$ elements. Without loss of generality, let focus on its first slice over the first axis:*

$$P_1^a = \begin{bmatrix} p_{1,1,1}^a & p_{1,1,2}^a & p_{1,1,3}^a & p_{1,1,4}^a \\ p_{1,2,1}^a & p_{1,2,2}^a & p_{1,2,3}^a & p_{1,2,4}^a \\ p_{1,3,1}^a & p_{1,3,2}^a & p_{1,3,3}^a & p_{1,3,4}^a \\ p_{1,4,1}^a & p_{1,4,2}^a & p_{1,4,3}^a & p_{1,4,4}^a \end{bmatrix}.$$

*If we analyze the model behavior on a relatively short time span (say, one year), then we can safely assume that the individual age is a static attribute. As such, those cortege changes that do not preserve the value of age are not available. To be more specific, cortege shifts*

$$z_1 \to z_2, z_1 \to z_4, z_2 \to z_1, z_2 \to z_3, z_3 \to z_2, z_3 \to z_4, z_4 \to z_1, z_4 \to z_3$$

*are prohibited. In this regard, we can rewrite $P_1^a$ by zeroing out the second and the fourth columns:*

$$P_1^a = \begin{bmatrix} p_{1,1,1}^a & 0 & p_{1,1,3}^a & 0 \\ p_{1,2,1}^a & 0 & p_{1,2,3}^a & 0 \\ p_{1,3,1}^a & 0 & p_{1,3,3}^a & 0 \\ p_{1,4,1}^a & 0 & p_{1,4,3}^a & 0 \end{bmatrix}.$$

*Analogously, in the matrix $P_1^a$ the second and fourth columns are zero. Instead, in the matrices $P_2^a$ and $P_4^a$ the first and third columns are zero*

Acting in a similar fashion, one can handle the situation when there are two types of agents in the system – ordinary agents (native users) and stubborn ones (online bots). To do this, one needs only to introduce one additional attribute with two possible values, representing both the agent types. The resulting transition matrix will be augmented to explain how these two types of agents should act in influence events.

*C. Introducing Ranking Algorithms*

Ranking algorithms – the mechanisms embedded in the architecture of online platforms that sort the information users receive to ensure that the most important content will not pass away – are the signature of the modern online networks. There are many types of ranking algorithms: those that prioritize the information in the news feed, those that sort the comments attached to posts, those that recommend new friends or followees to users (as these accounts may include some interesting information), etc. Formally speaking, the last example relates more to so-called recommendation systems, but for the sake of simplicity, I will not distinguish between these two terms.

Most ranking algorithms operate in a personalized way – the result of their work depends upon a given user's history of actions in the online network and, sometimes, on the Internet. Attempting to provide a user with the most valuable information, these algorithms may draw users into so-called filter bubbles [16].

In the seminal paper [11], the SCARDO model was equipped with the feature of ranking algorithms in an extremely simplified way: it was assumed that if the communicating agents' opinions are too distant (the distance between them is above a threshold $\Delta$), then with a probability $\delta$ the communication is prohibited. If the communication is allowed, then agents follow a standard opinion dynamics protocol and the recipient updates its opinion according to the transition matrix.

In this paper, I advance this approach by assuming that in the advanced SCARDO model, two selected agents communicate with a probability $\rho$, which is a function of the agents' characteristics $z_s$ and $z_l$:

$$\rho = f(z_s, z_l).$$

From the model perspective, to define a ranking algorithm, one needs only to set the function $f$. This function may encompass different facets of real ranking algorithms. For example, one can assume that the agents with close opinions or, say, close interests (for example, supporters of the same football team) have more chances to communicate because the ranking mechanism implies that such agents will likely enjoy talking to each other.

To define the function $f$, it would be convenient to use a square matrix $F = [f_{s,l}] \in \mathbb{R}^{M \times M}$, where $f_{s,l} = f(z_s, z_l)$. Let me illustrate my approach to the ranking algorithm operationalization. I will use the settings from Example 1.

*Example 2.* Let us assume that we consider the social system from Example 1 (where four possible cortege combinations exist) and the function $f$ is defined as follows:

$$F = \begin{bmatrix} 1 & 0.8 & 0.6 & 0.4 \\ 0.8 & 1 & 0.4 & 0.6 \\ 0.6 & 0.4 & 1 & 0.8 \\ 0.4 & 0.6 & 0.8 & 1 \end{bmatrix}.$$

From the matrix $F$ one can conclude that: (i) the more attributes two agents share, the more chances that they will be allowed to talk (note that according to $F$, agents with similar corteges always communicate, whereas the fully different agents – characterized by corteges $z_1$ and $z_4$ have interaction in four out of every ten cases); (ii) discrepancies in opinions have more effect on the communication probability than belonging to different age groups; (iii) the communication probability does not depend on who is a donor and who is a recipient – the matrix is symmetric (in the general case, one can employ an asymmetrical matrix whereby it is essential to emphasize the influence direction); (iv) the ranking mechanism has no preference for one opinion over the other – the matrix is symmetric with respect to the secondary diagonal.

It is also worth pointing out that the previous version of the ranking algorithm implementation (those that were applied in the seminal SCARDO model) is just a special case of the current approach.

## III. THE MEAN-FIELD APPROXIMATION FOR THE AUGMENTED MODEL

In this section, I derive the mean-field approximation for the advanced SCARDO model under the assumption that ranking algorithms interfere with communication processes. My purpose is to characterize the dynamics of macroscopic variables $y_1(t), \ldots, y_M(t)$ that represent the fractions of individuals with corteges $z_1, \ldots, z_M$ at time $t$ correspondingly.

First, I compute the probability that the population $Y_f(t)$ (not the fraction!) of agents with cortege $z_f$ will increase by one at this turn:

$$\Pr(Y_q(t+1) = Y_q(t) + 1) =$$

$$= \sum_{s=1}^{m} \frac{Y_s(t)}{N}(1-\delta_{s,q}) \sum_{l=1}^{m} \frac{Y_l(t)-\delta_{s,l}}{N} f_{s,l} \sum_{k=1}^{m} p^a_{s,l,k}\delta_{k,q},$$

where

$$\delta_{k,q} = \begin{cases} 0, & k \neq q \\ 1, & k = q \end{cases}$$

The probability that $Y_f(t)$ will decrease by one is obtained similarly:

$$\Pr(Y_q(t+1) = Y_q(t) - 1) =$$

$$= \sum_{s=1}^{m} \frac{Y_s(t)}{N}\delta_{s,q} \sum_{l=1}^{m} \frac{Y_l(t)-\delta_{s,l}}{N} f_{s,l} \sum_{k=1}^{m} p^a_{s,l,k}(1-\delta_{k,q}).$$

While deriving the last two expressions, I silently assumed that the agents communicate via a complete graph, so there is no structural bound in the model.

Given that we know $Y_q(t)$, we now can calculate the mathematical expectations of $Y_q(t+1)$:

$$\mathrm{E}[\,Y_q(t+1)] = Y_q(t) + \sum_{s,l,k} \frac{Y_s(t)}{N}\frac{Y_l(t)-\delta_{s,l}}{N} f_{s,l}p^a_{s,l,k}\Delta_{s,k,q},$$

where

$$\Delta_{s,k,q} = [(1-\delta_{s,q})\delta_{k,q} - \delta_{s,q}(1-\delta_{k,q})] = \delta_{k,q} - \delta_{s,q}.$$

For large $N$, we can safely write:

$$Y_q(t+1) = Y_q(t) + \sum_{s,l,k} \frac{Y_s(t)}{N}\frac{Y_l(t)}{N} f_{s,l}p^a_{s,l,k}\Delta_{s,k,q}.$$

After introducing the scaled time $\tau = t/N$ and the scaled time step $\delta\tau = 1/N$, one can compute:

$$\frac{y_q(\tau+\delta\tau) - y_q(\tau)}{\delta\tau} = \sum_{s,l,k} y_s(\tau)y_l(\tau)f_{s,l}p^a_{s,l,k}\Delta_{s,k,q}.$$

Note that in the previous formula, we pass from the absolute quantities $Y_s$ to the normalized ones $y_s$. Because $N$ is large, we can safely obtain:

$$\frac{dy_q(\tau)}{d\tau} = \sum_{s,l,k} y_s(\tau)y_l(\tau)f_{s,l}p^a_{s,l,k}\Delta_{s,k,q}. \quad (1)$$

I end this section by writing the following system of differential equations:

$$\begin{cases} \frac{dy_1(\tau)}{d\tau} = \sum_{s,l,k} y_s(\tau)y_l(\tau)f_{s,l}p^a_{s,l,k}\Delta_{s,k,1}, \\ \ldots \\ \frac{dy_M(\tau)}{d\tau} = \sum_{s,l,k} y_s(\tau)y_l(\tau)f_{s,l}p^a_{s,l,k}\Delta_{s,k,M} \end{cases} \quad (2)$$

and equipping it with the initial conditions

$$y_1(t_0) = y_1, \ldots, y_M(t_0) = y_M, \quad (3)$$

where $y_1, \ldots, y_M$ are nonnegative and sum to one. Note that from system (2) one can easily obtain the description of the opinion dynamics (at the macroscopic level) because the factions of opinion camps $x_1^1, \ldots, x_{m_1}^1$ (which I denote $y_1^o, \ldots, y_{m_1}^o$) can be obtained as

$$y_1^o = y_1 + \cdots + y_{\prod_{j=2}^{L} m_j}, \ldots, y_{m_1}^o =$$
$$= y_{(m_1-1)*\prod_{j=2}^{L} m_j} + \cdots + y_{m_1*\prod_{j=2}^{L} m_j}.$$

## IV. ANALYSIS OF THE CAUCHY PROBLEM (2), (3)

In this section, I attempt to answer the standard questions that may arise if one tries to approximate the behavior of a complex stochastic system with mean-field-based differential equations: (i) Does the Cauchy problem (2), (3) have a solution? (ii) Does this solution make physical sense (in our case, we expect that variables $y_1, \ldots, y_M$ should be nonnegative and sum to one because they are proportions that cover the entire population); (iii) Is this solution unique? (iv) Can the solution be extended to the whole axis (in fact, we are interested only in the right direction)?

First, it is straightforward to show that equation (1) can be rewritten as follows:

$$\frac{dy_q(\tau)}{d\tau} = \sum_{s,l} y_s(\tau) y_l(\tau) f_{s,l} p^a_{s,l,q} - y_q(\tau) \sum_l y_l(\tau) f_{q,l}. \quad (4)$$

Let compute the sum of all derivatives (4) (for brevity, I omit the argument $\tau$):

$$\sum_q \frac{dy_q}{d\tau} = \sum_q \sum_{s,l} y_s y_l f_{s,l} p^a_{s,l,q} - \sum_q y_q \sum_l y_l f_{q,l} =$$

$$= \sum_{s,l} y_s y_l f_{s,l} \underbrace{\sum_q p^a_{s,l,q}}_{=1} - \sum_{q,l} y_q y_l f_{q,l} = 0.$$

As such, the following statement is true.

**Statement 1.** *The quantity $u = y_1 + \cdots + y_M$ is the first integral of the autonomous system (2).*

Next, after making use of the Comparison theorem [17], one can easily prove that:

**Statement 2.** *The solution to the Cauchy problem (2), (2) (which exists and is unique due to the Picard–Lindelöf theorem) is a nonnegative function.*

Statements 1 and 2 demonstrate that the phase variables meet our intuition regarding what the proportions should look like – they should be nonnegative quantities, and because they cover all the population, their sum should be equal to one. Further, from these two statements, one can conclude that the solution to the Cauchy problem (2), (3) is a bounded function. Thus, the following result follows:

**Statement 3.** *The solution to the Cauchy problem (2), (3) can be extended to the whole axis.*

Furthermore, because the right side of the system (2) is a polynomial, we can guarantee that the solution to the Cauchy problem (2), (3) depends smoothly on the model parameters:

**Statement 4.** *The solution to the Cauchy problem (2), (3) is an analytic function of the model parameters $P^a, F$ and the initial condition.*

## V. Conclusion

I advanced the SCARDO model [11] to account for the heterogeneity of the influence processes in real societies. Previously, in the seminal paper, it was assumed that the outcome probability distribution was a function of the interacting agents' opinion only. I modified this assumption by augmenting the parameter space with individual non-opinion attributes and postulating that the outcome of the influence event is a function of all the user attributes, not only opinions. Mathematically, the key ingredient of the SCARDO model – the transition matrix that prescribes the outcomes of all possible pairwise interactions in terms of probability distributions – was augmented to incorporate information on how agents' non-opinion characteristics affect opinion dynamics. Next, I equipped the advanced SCARDO model with a ranking algorithm presented as a function that, for a given pair of agents, generates the probability that they will communicate, based on the agents' attributes. For the resulting model, a mean-field approximation was elaborated that describes the behavior of the model at the macroscopic level via an autonomous system of ordinary differential equations under the assumptions that the number of agents is large and the agents communicate via a complete social graph. In this differential equation system, the rate of change of the agent population with a given combination of attributes is a nonlinear function of other agent populations weighted with the augmented transition matrix components and the elements of the matrix that represent the ranking algorithm. The properties of this system (equipped with an initial condition) were thoroughly studied. I proved that the resulting Cauchy problem has a unique solution which is an analytic function of the model parameters. This solution is a bounded function that could be extended to the whole axis.

The next step is to apply the model elaborated in this paper to longitudinal data that includes information on individuals' non-opinion characteristics. Such data can be retrieved, for example, from online social networks [18], [19]. The interesting question to answer is whether accounting for individuals' non-opinion attributes can lead to different predictions at both the microscopic and macroscopic levels.

### Acknowledgment

The author is grateful to anonymous reviewers for their invaluable comments.